\documentclass[%
 reprint,
 amsmath,amssymb,
 aps,
]{revtex4-2}

\usepackage{graphicx}
\usepackage{dcolumn}
\usepackage{bm}
\usepackage{amsmath}
\usepackage[utf8]{inputenc}
\usepackage{textgreek}

\begin{document}

\preprint{APS/123-QED}

\title{Improved Threshold for Particle-Induced Magnetic Avalanche in Single-Molecule Magnets Using Fe$_8$ Molecule}

\author{Bailey Kohn}
\email{b.pickard17@tamu.edu}
\affiliation{Department of Physics and Astronomy, Texas A\&M University, College Station, TX, USA}
\author{Rupak Mahapatra}
\affiliation{Department of Physics and Astronomy, Texas A\&M University, College Station, TX, USA}
\author{Ivan Borzenets}
\affiliation{Department of Physics and Astronomy, Texas A\&M University, College Station, TX, USA}
\author{Tom Melia}
\affiliation{Kavli IPMU (WPI), UTIAS, The University of Tokyo, Kashiwa, Chiba 277-8583, Japan} 
\author{Michael Nippe}
\affiliation{Department of Chemistry, Texas A\&M University, College Station, TX, USA}
\author{Lok Raj Pant}
\affiliation{Department of Physics and Astronomy, Texas A\&M University, College Station, TX, USA}
\author{Surjeet Rajendran}
\affiliation{Department of Physics \& Astronomy, The Johns Hopkins University, Baltimore, Maryland 21218}
\author{Anna Schmautz}
\affiliation{Department of Chemistry, Texas A\&M University, College Station, TX, USA}
\author{Amis Sharma}
\affiliation{Department of Physics and Astronomy, Texas A\&M University, College Station, TX, USA}

\date{\today}

\begin{abstract}
Extending the original work on the development of a magnetic avalanche detector using Mn$_{12}$-ac single-molecule magnet (SMM), we report the results on a significantly lower threshold magnetic avalanche detector using Fe$_8$ SMM. Fe$_8$ has an order of magnitude smaller relaxation time that is expected to produce at least 3 orders of magnitude lower avalanche threshold compared to Mn$_{12}$-ac. We confirm this experimentally through the detection of gamma particles with energy at least two orders of magnitude lower than the original Mn$_{12}$-ac detection demonstrated using alpha particles, limited by the experimentally available radiation source. The true threshold of avalanche may be significantly lower and will be explored with lower energy x-rays and potentially infrared photons.
\end{abstract}

\maketitle

Detecting a single quanta of low energy is of particular interest for the scientific community, whether infra-red photons for quantum computing or sub-GeV dark matter particles or dark photons in the search for dark matter. The detection of low energies is often accomplished through signal amplification of the initial energy deposition. Recently, it was shown that a signal due to an alpha particle interaction (MeV) can be amplified in the single-molecule magnet (SMM) Mn$_{12}$-ac by the phenomenon known as magnetic avalanche in a manner analogous to a bubble chamber\cite{Kohn_Mn12, traditional_bubble_chamber}. It had previously been theorized that a signal as low as 10 meV could be detected by this method using an SMM with the right parameters \cite{Bunting_magnetic_bubble_chamber}. In this work, we take the next step towards this goal by 1) repeating the method of signal amplification in a second SMM, Fe$_8$, and 2) detecting signal from a gamma particle (keV) thus improving the energy threshold by 3 orders of magnitude. 

\begin{figure}[!htbp]
    \centering
    \includegraphics[width=\linewidth]{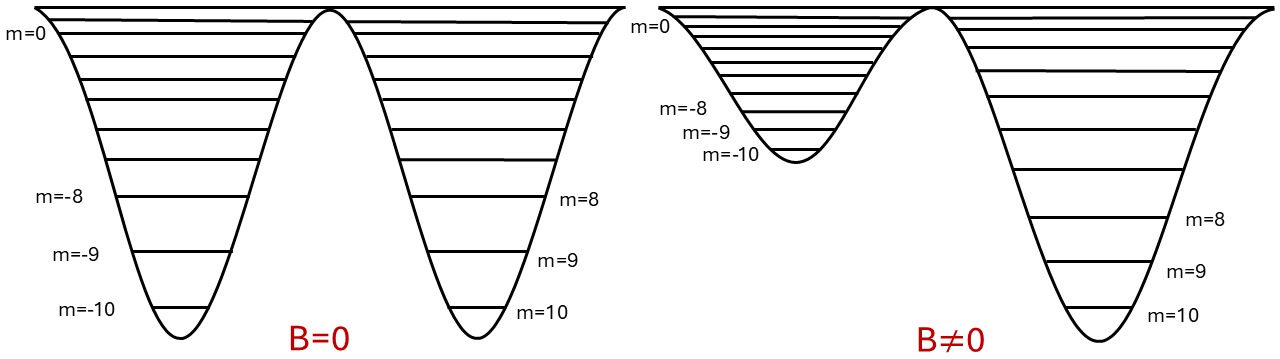}
    \caption{The spin potential of an SMM. Left: In the absence of an external field the up and down states have equal probability of being occupied. Right: In the presence of a magnetic field the potential shifts, forming a ground state in alignment with the field and an anti-aligned meta-stable state.}
    \label{fig:potential_wells}
\end{figure}

Single-molecule magnets are high spin materials characterized by magnetic bistability leading to open magnetic hysteresis loops at low temperatures. They were discovered by Gatechi and Sessoli in 1993 when they studied the magnetic properties of a molecule \cite{Mn12_nature} synthesized by Lis \cite{Lis_Mn_structure} known as Mn$_{12}$-ac (Mn$_{12}$O$_{12}$(O$_2$CCH$_3$)$_{16}$(H$_2$O)${_4}$). Fe$_8$ ([(C$_6$H$_{15}$N$_3$)$_6$Fe$_8$O$_2$(OH)$_{12}$]Br$_7$(H$_2$O)Br$\cdot$8H$_2$O) was discovered soon after \cite{Barra_Fe8_original, Fe8_structure} as a second molecule in the new "single-molecule magnet" category. It was determined that these materials have a spin potential that can be modeled as a double potential well (FIG. \ref{fig:potential_wells} left). The barrier height is dependent on the Zeeman energy and is given by
\begin{equation}
    \tilde{U}(B) = U -\frac{1}{2}\Delta E_{Zee}
\end{equation}
where $U$ is the barrier height intrinsic to the material and $\Delta E_{Zee}$ is the Zeeman splitting energy defined by
\begin{equation}
    \Delta E_{Zee} = 2\mu_Bg_JSB
\end{equation}
where $\mu_B$ is the Bohr magnetron, $g_J$ is the Land\'{e} g-factor, $S$ is the spin of the molecule, and $B$ is the external magnetic field \cite{Bunting_magnetic_bubble_chamber}. The second term in the barrier height causes the magnetic potential to shift when the easy axis of the molecule is aligned with a magnetic field (FIG \ref{fig:potential_wells} right). The resulting anti-aligned metastable state is what causes an open hysteresis loop at low temperatures \cite{review_Friedman_2010}. Molecules can change their spin state either by gaining enough energy to climb over the barrier or by quantum tunneling. The time it takes for a spin to relax is dependent on barrier height, temperature, and other factors intrinsic to the material itself. For many SMMs, including Mn$_{12}$-ac and Fe$_8$, magnetic relaxation time can be modeled using the Arrhenius Law \cite{Mn12_nature, Barra_Fe8_original}, specifically
\begin{equation}
    \tau = \tau_0 \text{ exp}\left(\frac{\tilde{U}(B)}{k_B T} \right)
\end{equation}
where $\tau_0$ is the relaxation time intrinsic to the molecule, $k_B$ is the Boltzmann constant, and $T$ is temperature. From the relaxation time, it is useful to define the blocking temperature, commonly defined as the temperature below which the magnetization is stable for 100 seconds. 

One specific phenomenon seen in SMMs as a consequence of their spin potential is magnetic avalanche. As an external field is applied, a significant fraction of molecules in an SMM crystal are trapped in the anti-aligned metastable spin state. If a small region of molecules through some means receives enough energy to overcome the barrier, they will relax into the ground state releasing the Zeeman energy difference between the two states to the surrounding molecules. This gives the neighboring molecules enough energy to flip states as well and repeat the process across the full crystal. This magnetic avalanche has been documented in many different types of SMMs by many different methods \cite{Avalanche_dueto_field_Suzuki,Paulsen1995_first_avalanche,Fe8_avalanche,Theory_deflagration,magnetic_deflageration_other} such as supplying surface acoustic waves \cite{Macia_avalanche_acoustic_waves, Hernandez_avalanche_wavetrigger}, heating one side of the crystal \cite{McHugh2007_avalanche_by_heat, Subedi_avalanche_heat}, or through particle interaction \cite{Kohn_Mn12, chen_avalanche_paper}. That an avalanche can be created by a particle interaction was first proposed by Bunting et al in 2017 \cite{Bunting_magnetic_bubble_chamber} and demonstrated by Chen et al in 2020 \cite{Chenthesis, chen_avalanche_paper} and confirmed by Kohn et al in 2026 \cite{Kohn_Mn12}. Bunting et al theorized that an SMM with the right parameters ($U = 50$K, $\tau_0 = 5\times10^{-12}$) would have an avalanche energy threshold as low as 10meV. A recent refinement of the calculation, however, has shown that this threshold is even more attainable, only requiring $\tau_0$ to be in the 0.1 ns range rather than in the ps range \cite{eberhardt_refined_sensitivity_calculation}. They also outline how this would serve as a dark photon detector and highlight several advantages.

In this work, we repeat our proof-of-concept work done with the molecule Mn$_{12}$-ac with a new molecule Fe$_8$. Mn$_{12}$-ac was the first SMM discovered and is one of the most studied of these molecules. Mn$_{12}$-ac has spin $S=10$, energy barrier $U=61$ K, and time constant $\tau_0 = 2.1\times10^{-7}$ s \cite{Mn12_nature}. For the magnetic relaxation to be stable for months, the blocking temperature calculated from equation (3) is 2K. Fe$_8$ is the second SMM discovered and it is also well studied. It also has spin $S=10$, energy barrier $U=24.5$ K, and time constant $\tau_0=3.4\times10^{-8}$ \cite{QMT_Fe8}. The blocking temperature for Fe$_8$ is lower, around 1K \cite{QMT_Fe8}. There are two main reasons for repeating our previous work. The first is verification that particle-induced avalanches can occur in SMMs besides Mn$_{12}$-ac. The second is to demonstrate a lower energy threshold. From the refined energy sensitivity calculation in ref. \cite{eberhardt_refined_sensitivity_calculation}, we can use
\begin{equation}
    E_{th}(B)\sim10^{-5}\beta^{5/2}\left[\frac{\alpha \tau_0}{\rho_s\Delta E_{Zee}(B)}\right]^{3/2}\left[\frac{\tilde{U}(B)}{k_B}\right]^{10}
\end{equation}
to estimate the energy threshold for an avalanche where $\beta$ is the specific heat assuming $E=\beta T^4V$, $\alpha$ is the thermal diffusivity, $\tau_0$ is the magnetic relaxation time given in equation (3), $\rho_s$ is the spin density defined as $\rho_s=1/V_{cell}$ where $V_{cell}$ is the unit cell volume, and $\tilde{U}(B)$ is the effective energy barrier given in equations (1) and (2). As mentioned before, Fe$_8$ has an order of magnitude lower $\tau_0$ than Mn$_{12}$-ac; and the thermal diffusivity for Fe$_8$ is also estimated to be lower by one to two orders of magnitude \cite{Fe8_avalanche}. We can use equation (4) to predict an energy threshold decrease of at least 3 orders of magnitude for a given external field, assuming all other factors are constant. We will test this using the same Am-241 sources as used in previous Mn$_{12}$-ac tests, which emit both 5 MeV alpha and 60 keV gamma particles. 

\begin{figure}
    \centering
    \includegraphics[width=\linewidth]{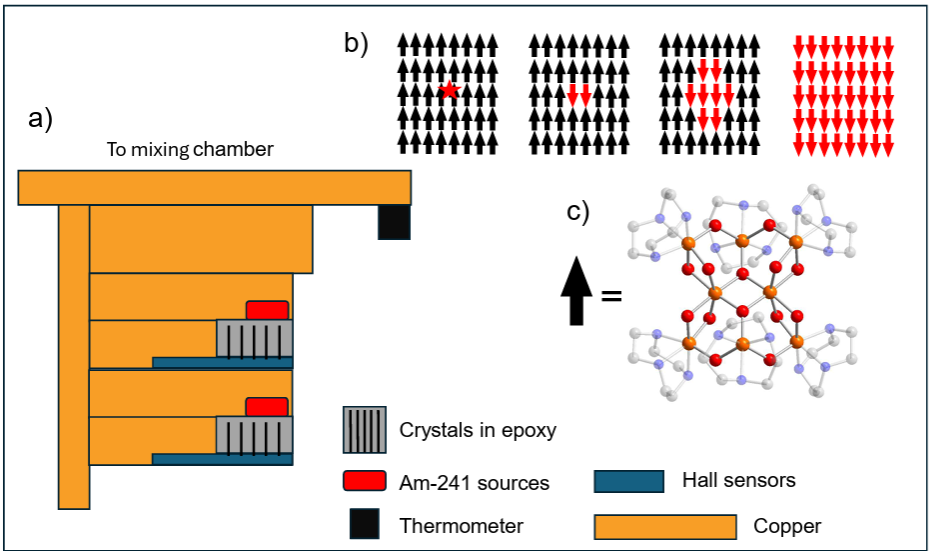}
    \caption{a) A diagram of the experimental set up. The lower crystals are fully covered by silver epoxy (covered sample) while there is a hole in the epoxy facing towards the Am-241 source in the upper crystal (open sample). A RuOx thermometer is mounted to the top of the cold finger leading to the fridge's mixing chamber. b) The propagation of an avalanche through a crystal. c) A molecular diagram of an Fe$_8$ molecule synthesized for this experiment with the counter-ion and hydrogen ions omitted for clarity.}
    \label{fig:setup}
\end{figure}

The Fe$_8$ crystals are synthesized following the method outlined by North \cite{Norththesischemistry}. Unlike Mn$_{12}$-ac, this material has two easy-axes and grows along both to make a flat diamond like shape. The largest of our crystals were roughly $1.26 \text{mm} \times 1.11\text{mm} \times 0.60\text{mm}$. We used silver epoxy to pack approximately 6 to 8 crystals each into two separate copper housings, being careful to align the crystals parallel such that one easy axis will align with the applied magnetic field. The silver epoxy acts as a heat sink and a barrier for alpha particles. Two Americium 241 (Am-241) sources were placed each facing a sample. One sample was fully covered with silver epoxy (called covered sample) fully blocking alpha particles, while one was only partially covered with epoxy, leaving the crystals exposed to alpha radiation (called open sample). Both samples were fitted with a Paragraf graphene hall sensor, separated by a layer of kapton tape. All of this, along with a resistive ruthenium oxide (RuOx) thermometer, were placed in an Oxford He dilution fridge on a cold finger that extended into a superconducting magnet. The dilution fridge has a base temperature of 8mK, but with both hall sensors on, the working base temperature for the experiment was around 80mK (see FIG. \ref{fig:setup}). We quickly discovered that our Fe$_8$ crystals shatter into powder when put under even a weak vacuum. The silver epoxy protects the covered sample, but the open sample's crystals would shatter in the dilution fridge's vacuum. To protect the crystals, we replaced the air in the vacuum chamber with nitrogen gas before beginning the cooling cycle. Around $-30^\circ$C, the crystals freeze allowing us to safely remove the nitrogen gas.

\begin{figure}
    \centering
    \includegraphics[width=\linewidth]{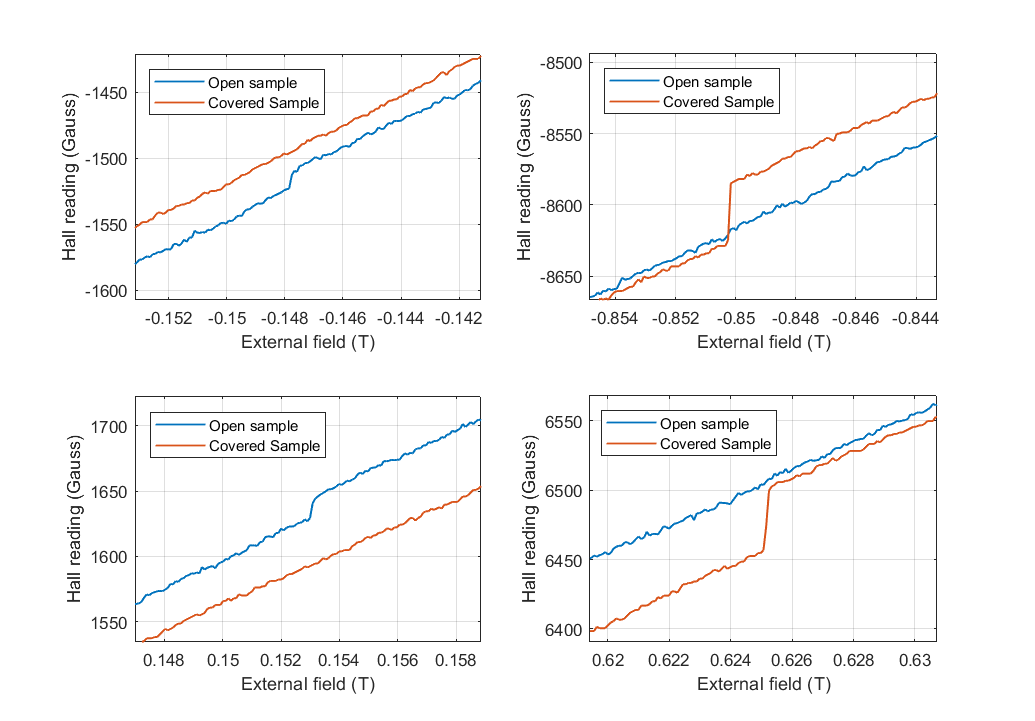}
    \caption{Magnetization vs. external field for a continuously ramping external field from $+1$ T to $-1$ T (top row) and back to $+1$ T (bottom row). The x-axis is the external field and the y-axis is the hall sensor data converted to Gauss for each sample. The four graphs are zoomed in to each avalanche observed; the top row corresponds to when the crystals were magnetized at +1 T and the field ramped to -1 T, while the bottom row is when the crystals were magnetized at -1 T and the field ramped to +1 T. Avalanches are seen in the open sample around $0.2$ T, consistent with those seen due to alpha particles in Mn$_{12}$-ac. Avalanches were also seen in the covered sample near $0.7$ T; these were induced by gamma radiation since all alpha radiation is blocked by epoxy in this sample.}
    \label{fig:continuous_ramp}
\end{figure}

Once at base temperature, we turn on the superconducting magnet to $\pm 1$ T to magnetize all molecules in alignment with the field. As we sweep the field towards the opposite orientation, the molecules stay aligned in their original magnetization. We first took data while continuously sweeping the external field at a rate of $0.04$ T/min. FIG. \ref{fig:continuous_ramp} shows the results from magnetizing at $+1$ T, sweeping to $-1$ T, then sweeping back to $+1$ T. Sharp, fast (order of 1 s) changes of magnetization of about 15 Gauss are seen in the open sample near $\pm0.2$ T. Comparing to our Mn$_{12}$-ac measurements, this is near the external field where we expect to see avalanches due to alpha particles. No avalanches were seen in the covered sample in this range either ramping up or down. Sharp changes in magnetization of about 40 Gauss are seen in the open sample around $\pm0.7$ T both ramping up and down. These were likely caused by gamma particles, since they can penetrate the silver epoxy covering the crystals. This process was repeated several times always with an avalanche in the open sample in the range of $\pm0.15$ to $\pm0.3$ T and an avalanche in the covered sample in the range from $\pm0.6$ to $\pm0.9$ T.  Each avalanche occurring in the lower field range was always on the order of 10 Gauss. Since alpha particles can only get to the crystals through a gap made in the epoxy, it is likely that only one or two crystals are actually affected. These may have been one of the smaller crystals, or misaligned with the field giving a smaller signal. The signal in the covered sample is larger always $\sim50$ Gauss, meaning the crystals are likely either larger or aligned to the field better.

We can use equation (4) to further verify that the avalanches in the open and covered samples were caused by alpha and gamma particles respectively. We take the values $V_{cell}=1.96\text{ nm}^3$ \cite{Pratt_Fe8_Vcell}, $\beta = 1.5\times10^{-6}$ eV/K$^4$/nm$^3$ (calculated from the Debye temperature $\theta_D = 19$ K \cite{Evangelisti_Fe8_specific_heat}) and $\alpha=2\times 10^{-6}$m$^2$/s \cite{Fe8_avalanche} along with the known values for $U$, $\tau_0$, and $S$ for Fe$_8$ mentioned previously. Taking a typical field value for an avalanche in each sample, we calculate 
\begin{equation}
    E_{th}(0.2\text{T})=5.5\text{ MeV}
\end{equation} 
and 
\begin{equation}
    E_{th}(0.7\text{T})=21\text{ keV},
\end{equation} 
each comparable to the alpha (5 MeV) and gamma (60 keV) energies of Am-241. 

\begin{figure}
    \centering
    \includegraphics[width=\linewidth]{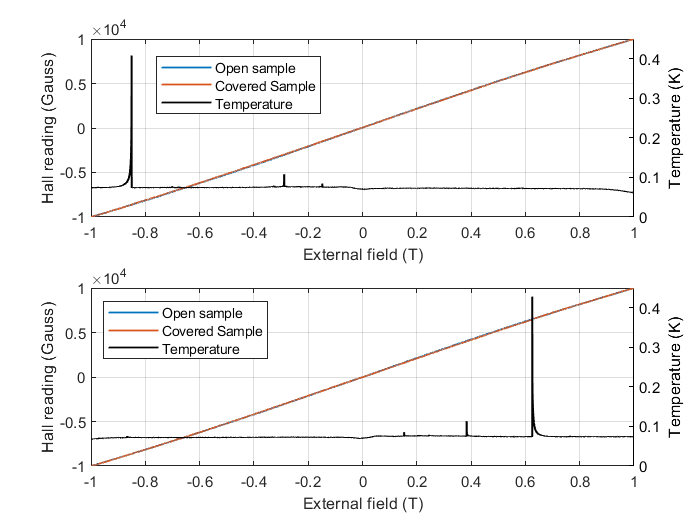}
    \caption{Data in FIG. \ref{fig:continuous_ramp} with temperature data overlaid. Data from when the crystals were magnetized at +1 T is shown in the top graph and data from when they were magnetized at -1 T is in the bottom graph. Temperature spikes are seen corresponding to the same external field as each avalanche in FIG. \ref{fig:continuous_ramp}. Two extra temperature spikes are seen near $-0.3$ T and $+0.4$ T which do not appear to correspond to any avalanche. These many have been caused by an avalanche in a crystal that was completely misaligned or too far away from the hall sensor to give a signal.}
    \label{fig:temperature_data}
\end{figure}

It is also enlightening to compare the hall sensor data to temperature data. When an avalanche occurs, the difference in energy from the metastable state to the ground state (equal to $\Delta E_{Zee}$) is released to the environment causing a rise in temperature. Since the Zeeman energy is directly proportional to the external field, we expect the temperature change to be proportional as well. FIG. \ref{fig:temperature_data} shows the temperature from the cold finger thermometer during the continuously ramping field run from FIG. \ref{fig:continuous_ramp}. A temperature spike proportional to the external field is clearly seen corresponding to each avalanche. Two extra spikes are seen that do not seem to correspond to an avalanche, however they are seen both ramping up and down and are proportional to the field, so they may have been caused by an avalanche in a crystal that had fallen so its easy axis was not in alignment with the hall sensor. The molecular density of Fe$_8$ is 1.918 g/cm$^3$ \cite{Fe8_structure}, so there are roughly $N=4\times10^{17}$ molecules in one of our crystals. Assuming the copper holder is 0.01 kg, and taking the specific heat of copper to be 0.1 J/kg K we can calculate the expected change in temperature read from the mounted thermometer for each avalanche to be:
\begin{equation}
    \Delta T (B=0.2T)=\frac{\Delta E_{Zee}N}{mc} \approx 30 \text{mK}
\end{equation}
\begin{equation}
    \Delta T (B=0.7T)=\frac{\Delta E_{Zee}N}{mc}  \approx 100\text{mK}.
\end{equation}
This is on the same order of magnitude as shown in the data, with deviations perhaps resulting from varying crystal sizes. However, the specific heat of copper is very dependent on temperature. At low temperature, the specific heat can be modeled by the sum of the electron and phonon contributions as shown in equation (9) \cite{Meissner}.
\begin{equation}
    C_{Cu} = c_{el} + c_{ph} = \gamma T + \frac{12 \pi^4}{5}R\left(\frac{T}{\Theta_D}\right)^3
\end{equation}
Where $T$ is the temperature, $\gamma$ is the electron specific heat coefficient, $R$ is the universal gas constant, and $\Theta_D$ is the Debye constant. Values for $\gamma$ and $\Theta_D$ were taken from ref. \cite{Specific_heat_lowT} to calculate the specific heat at $80$mK to be $~0.0008$ J/kg K, roughly 2 orders of magnitude lower than at room temperature. This results in a much higher temperature estimate than what is seen; however, there are other factors that have not been taken into account, such as the cooling power of the fridge at low temperature, the time delay of an avalanche, and the change in specific heat as the temperature rises. This exercise confirms that a large temperature spike is expected with an avalanche, especially at low temperatures. 

\begin{figure}
    \centering
    \includegraphics[width=\linewidth]{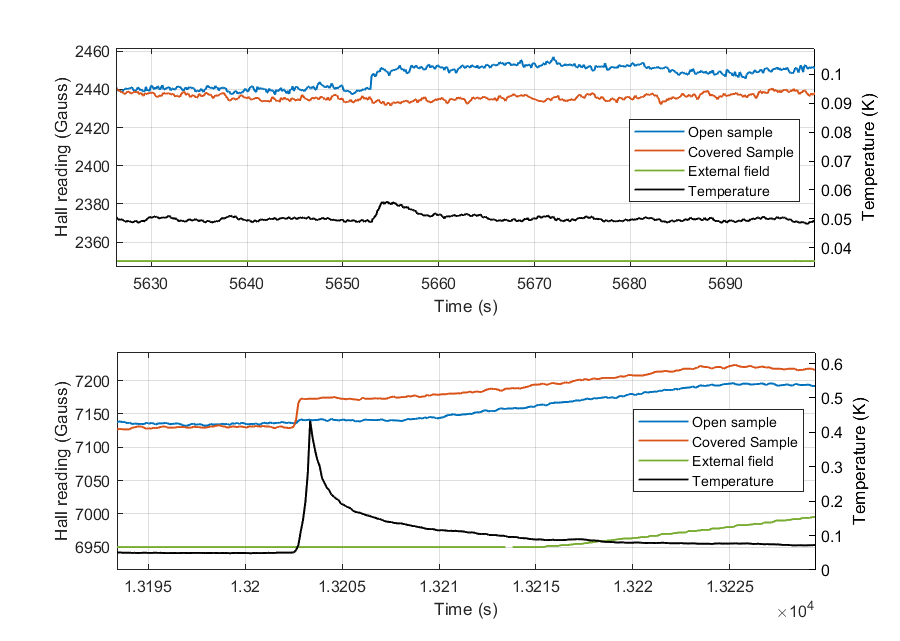}
    \caption{Magnetic field vs. time when the external field is held steady for 3 minutes at 50 Gauss intervals with temperature data. An avalanche is seen at 0.24 T in the open sample (top) and an avalanche is seen at 0.7 T in the covered sample (bottom). Temperature spikes are seen corresponding to both avalanches.}
    \label{fig:const_field}
\end{figure}

Next we needed to test that an avalanche can occur while the magnetic field is held constant. The crystals were magnetized at $-1.2$ T and the field slowly swept towards the opposite orientation as outlined above. Data was taken beginning at $+0.1$ T, where the field was then held constant. After 3 minutes, the field was raised by $50$ Gauss at a rate of $0.02$ T/min. We continued the process of waiting then raising the field until about $+0.3$ T, after which the alpha induced avalanche was likely to have already occurred. Since there is a distinct range of values where avalanches in the open and closed samples are observed, we then continuously ramped the field to $+0.6$ T and began 3 minute intervals of 50 Gauss again until the field reached $+1.0$ T. FIG. \ref{fig:const_field} shows these results. An avalanche of about 10 Gauss is seen in the open sample at 0.24 T and an avalanche of about 45 Gauss is seen at 0.71 Gauss in the covered sample. A large temperature spike is also seen coincident with each sample's avalanche of similar size to those seen previously. This measurement was taken 6 times, some saw no avalanches, some only saw one avalanche in either the open or the covered sample; FIG. \ref{fig:const_field} shows the only run that saw an avalanche in both samples in the same run. The crystals in the runs where no avalanches were seen may not have been fully magnetized in the beginning. For the runs where only one avalanche was seen, we may not have waited long enough to see the other avalanche. 

Based on these results, we can be sure that the features we have seen are the same particle induced avalanches previously observed in Mn$_{12}$-ac. Furthermore, these avalanches are seen to have been created by gamma radiation (60 keV), as well as alpha (5 MeV), demonstrating a three order of magnitude decrease in energy threshold from our previous work. It may possible to reach an even lower threshold as the external field is increased. The true energy threshold is only limited in our exploration by the availability of low-energy radioactive sources and will be explored in future work using either lower energy x-rays or tunable IR photons.

As highlighted in our previous work, an SMM based particle detector would have many advantages for low energy particle detection. First, avalanches are contained to one crystal, meaning that run time can be extended with the addition of more crystals. With a sensitive read-out system, powdered crystals can be used to maximize runtime and detection probability. After many of the crystals have undergone avalanche, resetting the detector is as simple as reversing the external field's polarity. Furthermore, as seen by comparing this work to our previous work, different SMMs have different energy thresholds, therefore we can probe different energy ranges simply by using different molecules. Finally, we can use micro-fabrication techniques to create readout systems in very close proximity to the crystals, increasing resolution and sensitivity. 

\begin{acknowledgments}
This work was supported by the Strategic Transformative Research Program at Texas A\&M University (R.M. and M.N.). R.M. acknowledges DOE support through DE-SC0021051 award that was instrumental in providing equipment and facilities to carry out this experiment. M.N. is also supported by funding from the Welch foundation (A-1880). I. B. notes that this work was partly performed under NCRADA-NIWCPacific-25-466. T.M. was supported by the World Premier International Research Center Initiative (WPI) MEXT, Japan, and by JSPS KAKENHI grants JP18K13533, JP19H05810, JP20H01896 and JP20H00153. S.R.  was supported by the U.S.~Department of Energy~(DOE), Office of Science, National Quantum Information Science Research Centers, Superconducting Quantum Materials and Systems Center~(SQMS) under Contract No.~DE-AC02-07CH11359. S.R. is also supported in part by the U.S.~National Science Foundation (NSF) under Grant No. PHY-1818899 and the Simons Investigator Grant No. 827042. T.M. was supported by the World Premier International Research Center Initiative (WPI) MEXT, Japan, and by JSPS KAKENHI grants JP22K18712, JP24H00244, and JP26K07100.

\end{acknowledgments}

\bibliography{Fe8}

@article{review_Friedman_2010,
   author = "Friedman, Jonathan R. and Sarachik, Myriam P.",
   title = "Single-Molecule Nanomagnets", 
   journal= "Annual Review of Condensed Matter Physics",
   year = "2010",
   volume = "1",
   number = "Volume 1, 2010",
   pages = "109-128",
   doi = "https://doi.org/10.1146/annurev-conmatphys-070909-104053",
   url = "https://www.annualreviews.org/content/journals/10.1146/annurev-conmatphys-070909-104053",
   publisher = "Annual Reviews",
   issn = "1947-5462",
   type = "Journal Article",
  }

@Inbook{Paulsen1995_first_avalanche,
    author="Paulsen, C. and Park, J.-G.",
    editor="Gunther, Leon and Barbara, Bernard",
    title="Evidence for Quantum Tunneling of the Magnetization in Mn12A C",
    bookTitle="Quantum Tunneling of Magnetization --- QTM '94",
    year="1995",
    publisher="Springer Netherlands",
    address="Dordrecht",
    pages="189--207",
    isbn="978-94-011-0403-6",
    doi="10.1007/978-94-011-0403-6_11",
    url="https://doi.org/10.1007/978-94-011-0403-6_11"
}

@article{Avalanche_dueto_field_Suzuki,
  title = {Propagation of Avalanches in ${\mathrm{Mn}}_{12}$-Acetate: Magnetic Deflagration},
  author = {Suzuki, Yoko and Sarachik, M. P. and Chudnovsky, E. M. and McHugh, S. and Gonzalez-Rubio, R. and Avraham, Nurit and Myasoedov, Y. and Zeldov, E. and Shtrikman, H. and Chakov, N. E. and Christou, G.},
  journal = {Phys. Rev. Lett.},
  volume = {95},
  issue = {14},
  pages = {147201},
  numpages = {4},
  year = {2005},
  month = {Sep},
  publisher = {American Physical Society},
  doi = {10.1103/PhysRevLett.95.147201},
  url = {https://link.aps.org/doi/10.1103/PhysRevLett.95.147201}
}

@article{Fe8_avalanche,
  title = {Quantum ignition of deflagration in the ${\mathrm{Fe}}_{8}$ molecular magnet},
  author = {Leviant, Tom and Keren, Amit and Zeldov, Eli and Myasoedov, Yuri},
  journal = {Phys. Rev. B},
  volume = {90},
  issue = {13},
  pages = {134405},
  numpages = {6},
  year = {2014},
  month = {Oct},
  publisher = {American Physical Society},
  doi = {10.1103/PhysRevB.90.134405},
  url = {https://link.aps.org/doi/10.1103/PhysRevB.90.134405}
}

@article{Theory_deflagration,
  title = {Theory of magnetic deflagration in crystals of molecular magnets},
  author = {Garanin, D. A. and Chudnovsky, E. M.},
  journal = {Phys. Rev. B},
  volume = {76},
  issue = {5},
  pages = {054410},
  numpages = {17},
  year = {2007},
  month = {Aug},
  publisher = {American Physical Society},
  doi = {10.1103/PhysRevB.76.054410},
  url = {https://link.aps.org/doi/10.1103/PhysRevB.76.054410}
}

@article{magnetic_deflageration_other,
  title = {Magnetic deflagration in ${\text{Gd}}_{5}{\text{Ge}}_{4}$},
  author = {Velez, S. and Hernandez, J. M. and Fernandez, A. and Maci\`a, F. and Magen, C. and Algarabel, P. A. and Tejada, J. and Chudnovsky, E. M.},
  journal = {Phys. Rev. B},
  volume = {81},
  issue = {6},
  pages = {064437},
  numpages = {7},
  year = {2010},
  month = {Feb},
  publisher = {American Physical Society},
  doi = {10.1103/PhysRevB.81.064437},
  url = {https://link.aps.org/doi/10.1103/PhysRevB.81.064437}
}

@article{QMT_Fe8,
  title = {Quantum Tunneling of the Magnetization in an Iron Cluster Nanomagnet},
  author = {Sangregorio, C. and Ohm, T. and Paulsen, C. and Sessoli, R. and Gatteschi, D.},
  journal = {Phys. Rev. Lett.},
  volume = {78},
  issue = {24},
  pages = {4645--4648},
  numpages = {0},
  year = {1997},
  month = {Jun},
  publisher = {American Physical Society},
  doi = {10.1103/PhysRevLett.78.4645},
  url = {https://link.aps.org/doi/10.1103/PhysRevLett.78.4645}
}

@phdthesis{Norththesischemistry,
    author = {Jeremy Micah North},
    title = {Synthesis and Characterization of Single-Molecule Magnets: Mn$_{12}$-Acetate, Fe$_8$Br$_8$, and Analogs},
    school = {The Florida State University},
    year = {2004}
}

@article{Fe8_structure,
    author = {Weighardt, Karl and Pohl, Klaus and Jibril, Ibrahim and Huttner, Gottfried},
    title = {Hydrolysis Products of the Monomeric Amine Complex (C6H15N3)FeCl3: The Structure of the Octameric Iron(III) Cation of {[(C6H15N3)6Fe8(μ3-O)2(μ2-OH)12]Br7(H2O)}Br·8H2O},
    journal = {Angewandte Chemie International Edition in English},
    volume = {23},
    number = {1},
    pages = {77-78},
    doi = {https://doi.org/10.1002/anie.198400771},
    url = {https://onlinelibrary.wiley.com/doi/abs/10.1002/anie.198400771},
    eprint = {https://onlinelibrary.wiley.com/doi/pdf/10.1002/anie.198400771},
    year = {1984}
}

@article{traditional_bubble_chamber,
  title = {Dark matter search results from the complete exposure of the PICO-60 ${\mathrm{C}}_{3}{\mathrm{F}}_{8}$ bubble chamber},
  author = {Amole, C. and Ardid, M. and Arnquist, I. J. and Asner, D. M. and Baxter, D. and Behnke, E. and Bressler, M. and Broerman, B. and Cao, G. and Chen, C. J. and Chowdhury, U. and Clark, K. and Collar, J. I. and Cooper, P. S. and Coutu, C. B. and Cowles, C. and Crisler, M. and Crowder, G. and Cruz-Venegas, N. A. and Dahl, C. E. and Das, M. and Fallows, S. and Farine, J. and Felis, I. and Filgas, R. and Girard, F. and Giroux, G. and Hall, J. and Hardy, C. and Harris, O. and Hillier, T. and Hoppe, E. W. and Jackson, C. M. and Jin, M. and Klopfenstein, L. and Kozynets, T. and Krauss, C. B. and Laurin, M. and Lawson, I. and Leblanc, A. and Levine, I. and Licciardi, C. and Lippincott, W. H. and Loer, B. and Mamedov, F. and Mitra, P. and Moore, C. and Nania, T. and Neilson, R. and Noble, A. J. and Oedekerk, P. and Ortega, A. and Piro, M.-C. and Plante, A. and Podviyanuk, R. and Priya, S. and Robinson, A. E. and Sahoo, S. and Scallon, O. and Seth, S. and Sonnenschein, A. and Starinski, N. and \ifmmode \check{S}\else \v{S}\fi{}tekl, I. and Sullivan, T. and Tardif, F. and V\'azquez-J\'auregui, E. and Walkowski, N. and Weima, E. and Wichoski, U. and Wierman, K. and Yan, Y. and Zacek, V. and Zhang, J.},
  collaboration = {PICO Collaboration},
  journal = {Phys. Rev. D},
  volume = {100},
  issue = {2},
  pages = {022001},
  numpages = {9},
  year = {2019},
  month = {Jul},
  publisher = {American Physical Society},
  doi = {10.1103/PhysRevD.100.022001},
  url = {https://link.aps.org/doi/10.1103/PhysRevD.100.022001}
}

@article{Kohn_Mn12,
   title={Particle detection using magnetic avalanches in single-molecule magnet crystals},
   ISSN={2470-0029},
   url={http://dx.doi.org/10.1103/r1x8-qc2f},
   DOI={10.1103/r1x8-qc2f},
   journal={Physical Review D},
   publisher={American Physical Society (APS)},
   author={Kohn, B. and Chen, H. and Mahapatra, R and Agnolet, G. and Borzenets, I. and Bunting, P. and Long, J. and Lu, M. and Melia, T. and Nippe, M. and Pant, L.R. and Rajendran, S. and Schmautz, A. and Sherma, A.},
   year={2026},
   month={May} 
}

@Article{Mn12_nature,
	author={Sessoli, R.
	and Gatteschi, D.
	and Caneschi, A.
	and Novak, M. A.},
	title={Magnetic bistability in a metal-ion cluster},
	journal={Nature},
	year={1993},
	volume={365},
	number={6442},
	pages={141-143},
	issn={1476-4687},
	doi={10.1038/365141a0},
	url={https://doi.org/10.1038/365141a0}
}

@article{Barra_Fe8_original,
    doi = {10.1209/epl/i1996-00544-3},
    url = {https://doi.org/10.1209/epl/i1996-00544-3},
    year = {1996},
    month = {jul},
    publisher = {},
    volume = {35},
    number = {2},
    pages = {133},
    author = {A.-L. Barra and P. Debrunner and D. Gatteschi and Ch. E. Schulz and R. Sessoli},
    title = {Superparamagnetic-like behavior in an octanuclear iron cluster},
    journal = {Europhysics Letters}
}

@article{Bunting_magnetic_bubble_chamber,
    title = {Magnetic bubble chambers and sub-GeV dark matter direct detection},
    author = {Bunting, Philip C. and Gratta, Giorgio and Melia,  Tom and Rajendran, Surjeet},
    journal = {Phys. Rev. D},
    volume = {95},
    issue = {9},
    pages = {095001},
    numpages = {12},
    year = {2017},
    month = {May},
    publisher = {American Physical Society},
    doi = {10.1103/PhysRevD.95.095001},
    url = {https://link.aps.org/doi/10.1103/PhysRevD.95.095001}
}

@article{Macia_avalanche_acoustic_waves,
    title = {Observation of phonon-induced magnetic deflagration in manganites},
    author = {Maci\`a, F. and Hern\'andez-M\'{\i}nguez, A. and Abril, G. and Hernandez, J. M. and Garc\'{\i}a-Santiago, A. and Tejada, J. and Parisi, F. and Santos, P. V.},
    journal = {Phys. Rev. B},
    volume = {76},
    issue = {17},
    pages = {174424},
    numpages = {8},
    year = {2007},
    month = {Nov},
    publisher = {American Physical Society},
    doi = {10.1103/PhysRevB.76.174424},
    url = {https://link.aps.org/doi/10.1103/PhysRevB.76.174424}
}

@article{Subedi_avalanche_heat,
  title = {Onset of a Propagating Self-Sustained Spin Reversal Front in a Magnetic System},
  author = {Subedi, P. and V\'elez, S. and Maci\`a, F. and Li, S. and Sarachik, M. P. and Tejada, J. and Mukherjee, S. and Christou, G. and Kent, A. D.},
  journal = {Phys. Rev. Lett.},
  volume = {110},
  issue = {20},
  pages = {207203},
  numpages = {5},
  year = {2013},
  month = {May},
  publisher = {American Physical Society},
  doi = {10.1103/PhysRevLett.110.207203},
  url = {https://link.aps.org/doi/10.1103/PhysRevLett.110.207203}
}

@article{McHugh2007_avalanche_by_heat,
	title = {Effect of quantum tunneling on the ignition and propagation of magnetic avalanches in ${\mathrm{Mn}}_{12}$ acetate},
	author = {McHugh, S. and Jaafar, R. and Sarachik, M. P. and Myasoedov, Y. and Finkler, A. and Shtrikman, H. and Zeldov, E. and Bagai, R. and Christou, G.},
	journal = {Phys. Rev. B},
	volume = {76},
	issue = {17},
	pages = {172410},
	numpages = {4},
	year = {2007},
	month = {Nov},
	publisher = {American Physical Society},
	doi = {10.1103/PhysRevB.76.172410},
	url = {https://link.aps.org/doi/10.1103/PhysRevB.76.172410}
}

@article{Hernandez_avalanche_wavetrigger,
  title = {Quantum Magnetic Deflagration in ${\mathrm{Mn}}_{12}$ Acetate},
  author = {Hern\'andez-M\'{\i}nguez, A. and Hernandez, J. M. and Maci\`a, F. and Garc\'{\i}a-Santiago, A. and Tejada, J. and Santos, P. V.},
  journal = {Phys. Rev. Lett.},
  volume = {95},
  issue = {21},
  pages = {217205},
  numpages = {4},
  year = {2005},
  month = {Nov},
  publisher = {American Physical Society},
  doi = {10.1103/PhysRevLett.95.217205},
  url = {https://link.aps.org/doi/10.1103/PhysRevLett.95.217205}
}

@article{Lis_Mn_structure,
    author = "Lis, T.",
    title = "{Preparation, structure, and magnetic properties of a dodecanuclear mixed-valence manganese carboxylate}",
    journal = "Acta Crystallographica Section B",
    year = "1980",
    volume = "36",
    number = "9",
    pages = "2042--2046",
    month = "Sep",
    doi = {10.1107/S0567740880007893},
    url = {https://doi.org/10.1107/S0567740880007893},
}

@article{Specific_heat_lowT,
    title = {Specific Heat of Copper, Silver, and Gold below 30\ifmmode^\circ\else\textdegree\fi{}K},
    author = {Martin, Douglas L.},
    journal = {Phys. Rev. B},
    volume = {8},
    issue = {12},
    pages = {5357--5360},
    numpages = {0},
    year = {1973},
    month = {Dec},
    publisher = {American Physical Society},
    doi = {10.1103/PhysRevB.8.5357},
    url = {https://link.aps.org/doi/10.1103/PhysRevB.8.5357}
}

@ARTICLE{Meissner,
    author={Pant, L. R. and Sharma, A. and Mirabolfathi, N. and Mahapatra, R. and Borzenets, I. V. and Finnegan, P. S. and Arrington, C. L. and John, C. St and Carr, S. M.},
    journal={IEEE Transactions on Applied Superconductivity}, 
    title={Cryogenic Microcalorimeter Prototype Using the Magnetic Superconducting Transition}, 
    year={2025},
    volume={35},
    number={5},
    pages={1-5},
    doi={10.1109/TASC.2025.3540729}
}

@misc{eberhardt_refined_sensitivity_calculation,
      title={Refined Sensitivity Estimates for Single-Molecule Magnet Dark Matter Detectors}, 
      author={Andrew Eberhardt and Tomoya Fukui and Ryosuke Takehara and Ryotaro Ohno and Yuta Mizukami and Kenichiro Hashimoto and Takanori Fukushima and Shigeki Matsumoto and Tom Melia and Kouki Nozaki and Surjeet Rajendran},
      year={2026},
      eprint={2607.01868},
      archivePrefix={arXiv},
      primaryClass={hep-ph},
      url={https://arxiv.org/abs/2607.01868}, 
}

@article{Pratt_Fe8_Vcell,
    title = {Dipolar ordering in a molecular nanomagnet detected using muon spin relaxation},
    author = {Pratt, F. L. and Micotti, E. and Carretta, P. and Lascialfari, A. and Arosio, P. and Lancaster, T. and Blundell, S. J. and Powell, A. K.},
    journal = {Phys. Rev. B},
    volume = {89},
    issue = {14},
    pages = {144420},
    numpages = {7},
    year = {2014},
    month = {Apr},
    publisher = {American Physical Society},
    doi = {10.1103/PhysRevB.89.144420},
    url = {https://link.aps.org/doi/10.1103/PhysRevB.89.144420}
}

@article{Evangelisti_Fe8_specific_heat,
  title = {Giant Isotope Effect in the Incoherent Tunneling Specific Heat of the Molecular Nanomagnet ${\mathrm{Fe}}_{8}$},
  author = {Evangelisti, M. and Luis, F. and Mettes, F. L. and Sessoli, R. and de Jongh, L. J.},
  journal = {Phys. Rev. Lett.},
  volume = {95},
  issue = {22},
  pages = {227206},
  numpages = {4},
  year = {2005},
  month = {Nov},
  publisher = {American Physical Society},
  doi = {10.1103/PhysRevLett.95.227206},
  url = {https://link.aps.org/doi/10.1103/PhysRevLett.95.227206}
}

@misc{chen_avalanche_paper,
      title={Quantum Detection using Magnetic Avalanches in Single-Molecule Magnets}, 
      author={Hao Chen and Rupak Mahapatra and Glenn Agnolet and Michael Nippe and Minjie Lu and Philip C. Bunting and Tom Melia and Surjeet Rajendran and Giorgio Gratta and Jeffrey Long},
      year={2020},
      eprint={2002.09409},
      archivePrefix={arXiv},
      primaryClass={physics.ins-det},
      url={https://arxiv.org/abs/2002.09409}, 
}

@phdthesis{Chenthesis,
    title    = {Magnetic Bubble Chamber Prototype Development},
    school   = {Texas A\&M University},
    author   = {Chen, Hao},
    year     = {2019}, 
    orcid    = {0000-0002-8395-8526}
}

\end{document}